\documentclass[reprint,aps]{revtex4}
\usepackage{amssymb,epstopdf,amsmath,enumerate,color}
\usepackage{subfigure}
\usepackage{graphicx}
\usepackage{float}

\newcommand{\Array}[2]{\left(\begin{array}{#1}#2\end{array}\right)}

\begin{document}
	
	\title{A Factorized Mass Structure of Fermions and Its Fit}    
	\author{Qingfeng Cao$^1$\footnote{E-mail:252721241562cao@stu.xjtu.edu.cn.}, Ying Zhang$^{1,2}$\footnote{Corresponding author. E-mail:hepzhy@mail.xjtu.edu.cn.}}
	\address{$^1$School of Science, Xi'an Jiaotong University, Xi'an, 710049, China\\
		$^2$ Institute of Modern Physics, Xi'an Jiaotong University, Xi'an, 710049, China}

	\begin{abstract}
		The structure of the mass matrix, a challenging problem in the Standard Model, is closely related to flavor phenomenology and the understanding of the Yukawa interaction. 
		We derive a factorized mass structure based on observed fermion mass hierarchies, investigating the role of $SO(2)^f$ family symmetry in explaining the approximate degeneracy of light quark generations and its connection to flavor mixing. 
		Our calculation includes the slightest modification from the lightest fermions by $\mathcal{O}(h^2)$ hierarchy corrections. 
		Using this model-independent framework, we systematically analyze all mass patterns and perform comprehensive fits to both quark CKM and lepton PMNS mixing data with extended normal-ordered Dirac neutrinos. 
		The results demonstrate this framework's capacity to unify flavor phenomena while simultaneously providing new insights into the fundamental nature of Yukawa interactions.
		

		
	\end{abstract}
	
	\maketitle

	\section{Motivation}
	\label{sec.intro}
	Since its establishment in the 1960s, the Standard Model (SM) has been extensively validated by high energy physics experiments. 
	In addition to describing strong and electroweak gauge interactions, the SM incorporates Yukawa interactions between the Higgs field and fermions. The Yukawa interactions determine the masses of quarks, charged leptons, and even neutrinos (if neutrinos are Dirac fermions and Yukawa terms generate their masses), while also governing flavor-changing processes through the quark Cabibbo-Kobayashi-Maskawa (CKM) mixing matrix and the lepton Pontecorvo-Maki-Nakagawa-Sakata (PMNS) mixing matrix. 
	In the SM, Yukawa couplings are introduced as family-dependent complex parameters without deeper theoretical justification. While this parameterization successfully accommodates observed phenomena, it fails to provide any fundamental explanation for either the hierarchical structure or specific values of these couplings \cite{ZupanCERN2019,FeruglioEPJC2015}. 
	This theoretical shortcoming leaves several profound questions about flavor physics unresolved, including the dynamical origin of fermion hierarchical masses, the fundamental source of CP violation, and the striking contrast between quark mixing and lepton mixing. 
	The absence of predictive principles governing these Yukawa couplings represents one of the most significant unresolved challenges in the SM.
	Numerous flavor models have been proposed, motivated by theoretical frameworks \cite{DimouPRD2016,TooropJHEP2010}, discrete symmetries \cite{SFKingRPP2013,YamanakaPRD}, or mixing patterns \cite{HarrisonPLB2002}. Yet, increasingly precise experimental data, particularly measurements of mixing parameters, pose significant challenges to these models. 
	
	What defines a desired flavor structure? 
	It should satisfy three fundamental requirements:
	(1) It must eliminate parameter redundancies. 
	Consider the quark sector as an example. To account for CP violation in the CKM matrix, the up-type quark mass matrices, $M^{u,d}$, must generally be complex, introducing $3\times 3\times2=18$ degrees of freedom (d.o.f.) for each matrix. 
	However, flavor observables consist of only 6 quark masses, 3 mixing angles, and 1 CP-violating phase, totaling 10 parameters, far fewer than the 36 d.o.f. in the Yukawa couplings. 
	This mismatch implies that a faithful flavor structure must describe quark/lepton flavor with 10 independent parameters.
	(2) The theory must consistently reproduce all experimental observables, including fermion masses, mixing angles, and CP-violating phases. This requirement imposes crucial constraints on theoretical approaches: in the quark sector, we cannot simply adopt the up-type quark diagonal basis while focusing exclusively on mixing effects from down-type quark diagonalization, as this would preserve the up-type mass hierarchy without explaining its origin. Similarly, for leptons, the PMNS mixing pattern cannot be adequately explained by considering only the neutrino mass matrix while neglecting the mass hierarchy of charged leptons. A complete description must simultaneously account for both mass hierarchies and mixing patterns in each sector through a unified treatment of their respective mass matrices.
	(3) The theory should establish an organized structure for Yukawa couplings that transcends mere phenomenological fitting, analogous to how gauge interactions derive from $SU(3)_C \times SU(2)_L\times U(1)_Y$. The similarity between quark and lepton mixing structure suggests that such a structure may originate from unified principles, whether through common mass matrix forms or shared symmetry breaking mechanisms, that simultaneously govern both sectors.

	Recent work has explored factorized mass matrices motivated by the hierarchical structure of fermion masses \cite{Zhang2025arxiv}. In this framework, the quark mass matrix $M^q$  (for $q=u,d$)  is factorized into a product of a diagonal phase matrix and a real symmetric matrix characterized by two free parameters, $l_1$ and $l_2$.
	Unlike top-down approaches that impose theoretical assumptions a priori, this approach adopts a bottom-up route, constructing the flavor structure solely from the general phenomenological characteristics of quarks and charged leptons. 
	This parameterization offers model-independent insights into the origin of CP violation and the interplay between fermion masses and mixing. Crucially, it reorganizes the SM Yukawa couplings into a minimalistic structure free of redundant d.o.f.. 
	Preliminary studies of a flat pattern ($l_1=l_2=1$) in both quark and lepton sectors show promising results \cite{Zhang2025arxiv}. 
	However, a comprehensive exploration of alternative parameterizations is essential to fully map the landscape of flavor dynamics.
	
	In this paper, we systematically investigate all mass patterns spanned by the parameters $l_1$ and $l_2$ using an updated computational framework. 
	The quark and lepton masses are treated without rescaling for mass-scale effects. This choice is justified because flavor mixing exhibits negligible sensitivity to absolute mass hierarchies, which contribute only perturbative corrections. 
	The paper is organized as follows: Sec. \ref{sec.MassPattern} reviews the factorized mass matrix formalism proposed in \cite{Zhang2025arxiv}.
	Sec. \ref{sec.AllPatterns} examines connections between this general parameterization and some popular mass patterns in literature.
	Flavor mixing matrices are derived in Sec. \ref{sec.mixing}, preceded by critical discussions on mass basis selection and implications of  $SO(2)$ family symmetry breaking.
	Sec. \ref{sec.fit} presents fit results for both quark and lepton sectors, including a discussion on neutrino nature and its mass ordering.
	A concise summary concludes the paper.

	\section{The factorized structure of the Mass matrix}
	\label{sec.MassPattern}
	
	\subsection{Mass Matrix Reconstruction in the hierarchy limit}
	The quark mass matrix $M^q$ ($q=u,d$) is diagonalized into physical eigenvalues by bi-unitary transformations $U_L^q$ and $U_R^q$:
	\begin{eqnarray}
		U_L^qM^q(U_R^q)^\dag=\text{ diag}(m_1^q,m_2^q,m_3^q). 
		\label{eq.UMU0001}
	\end{eqnarray}
	In the quark mixing, only left-handed $U_L^q$, rather than right-handed $U_R^q$, contributes to the CKM mixing $U_{CKM}=U_L^u(U_L^d)^\dag$.
	For a random unitary matrix $U'$, $M^qU'$ shares identical masses with $M^q$, implying right-handed transformation is non-unique for quark physical masses. 
	To eliminate unphysical $U_R^q$, we adopt the unitary condition $U_R^q=U_L^q$,  in which $M^q$ becomes a hermitian matrix \cite{RamondNPB1993}. (Alternatively, hermitian $M^q(M^q)^\dag$ can be adopted  to replace $M^q$,  which is diaogonalized by $U_L^q$ as 
	$U_L^qM^q(M^q)^\dag(U_L^q)^\dag=\textrm{diag}[(m_1^q)^2,(m_2^q)^2,(m_3^q)^2].$)

	Motivated by the hierarchical  masses of quarks
	$m_1^q\ll m_2^q\ll m_3^q$,	
	$M^q$ can be expanded  in power of mass hierarchy $h_{ij}^q=m_i^q/m_j^q$ ($i<j$):
	\begin{eqnarray}
		M^q=M_0^q+h_{23}^qM_1^q+h_{12}^qh_{23}^q M_{21}^q+(h_{23}^q)^2M_{22}^q+\mathcal{O}(h^3)
	\end{eqnarray}
	Here, $M_0^q$ is the leading order mass matrix, and $M_1^q,M_{2i}^q$ are order of $h^q_{ij}$ and $(h^q_{ij})^2$, respectively. 
When $M^q$ is normalized to the total quark mass $\sum_i m_i^q$, the effect of the lightest quark mass emerges at $\mathcal{O}(h^2)$ 
$$\frac{m_1^q}{\sum_im_i^q}\sim h_{12}^qh_{23}^q,$$ 
So, there is no the linear term in $h_{12}^q$.
	
	In the hierarchy limit  $h_{23}^q\rightarrow 0$, the lighter two families become degenerate. The matrix $M_0^q$ can be reconstructed solely from only the third row elements of $U_{L}^q$ 
	\begin{eqnarray}
		\frac{1}{{\sum m_i^q}}M_0^q\Big|_{ij}
		&=&(U_{L}^q)^*_{ik}\Array{ccc}{0 && \\ & 0 & \\ && 1}_{kl} U_{L,lj}^q
		\nonumber\\
		&=&(U_{L}^q)^*_{i3} U_{L,3j}^q
		\label{eq.U001U02}
	\end{eqnarray}
	Here, we adopt normalized mass matrix $M_0^q/{\sum m_i^q}$ to express the form after factoring out the total mass.
	Re-expressing $U^q_{L,3i}$ by 3 phases $\eta_i$ and 3 norms $l_i$ ($i=0,1,2$)
	\begin{eqnarray*}
		U^q_{L,33}=l_0e^{i\eta_0},~~~
		\frac{U^q_{L,31}}{U^q_{L,33}}=l_1e^{i\eta_1},~~~
		\frac{U^q_{L,32}}{U^q_{L,33}}=l_2e^{i\eta_2},
	\end{eqnarray*}
	Eq. (\ref{eq.U001U02}) can be written into a factorized form \cite{ZhangCJP2024}
	\begin{eqnarray}
		\frac{1}{\sum m_i^q}M_0^q
		=(K_L^q)^\dag
		M_N^q
		K_L^q.
		\label{eq.KMK2}
	\end{eqnarray}
	with the diagonal phase matrix $K_L^q$ and real symmetric matrix $M_N^q$
	\begin{eqnarray}
		K_L^q&=&\text{diag}\Big(e^{i\eta_1},e^{i\eta_2},1\Big)
		\\
		M_N^q&=&\frac{1}{l_1^2+l_2^2+1}\Array{ccc}{l_1^2 & l_1l_2 & l_1 \\
			l_1l_2 & l_2^2 & l_2 \\
			l_1 & l_2 & 1
		} .
		\label{eq.PatternM}
	\end{eqnarray}
	Parameter $l_0$ has been substituted by unitarity condition $l_0^2(l_1^2+l_2^2+1)=1$ in terms of  $U_{3i}U_{i3}^*=1$.
	The factorization separates the physical roles:
	$M_N^q$ determines the mass spectrum and $K_L^q$ introduces CP-violating phases for CKM mixing.
	
	For any flavor models, the mass matrix in the hierarchical limit must necessarily obey the following relations 
	\begin{eqnarray}
		&&\arg[M^q_{0,12}]+\arg[M^q_{0,23}]+\arg[M^q_{0,31}]=0,
		\label{Eq.rules1}\\
		&&\Big|\frac{M^q_{0,11}}{M^q_{0,21}}\Big|=\Big|\frac{M^q_{0,12}}{M^q_{0,22}}\Big|=\Big|\frac{M^q_{0,13}}{M^q_{0,23}}\Big|,
		\label{Eq.rules2}\\
		&&\Big|\frac{M^q_{0,11}}{M^q_{0,31}}\Big|=\Big|\frac{M^q_{0,12}}{M^q_{0,32}}\Big|=\Big|\frac{M^q_{0,13}}{M^q_{0,33}}\Big|.
		\label{Eq.rules3}
	\end{eqnarray}
	These constraints emerge naturally from the requirement that the model must reproduce hierarchical masses.
	
	\subsection{Hierarchy Correction}
	
	To accurately reproduce the physical masses of the first two families and achieve precise CKM mixing predictions, hierarchy corrections must be included.
	Any deviation from the exact relations in Eqs. (\ref{Eq.rules1}-\ref{Eq.rules3}) generates non-zero masses for the first and second generation fermions.
	Since the diagonal elements $M^q_{N,11}$ and $M^q_{N,22}$ can be used to define $l_1^2$ and $l_2^2$ respectively, the general perturbed form of $M_N^q$ is expressed as
	\begin{eqnarray}
		M_N^q=\frac{1}{l_1^2+l_2^2+1}\Array{ccc}{l_1^2 & l_1l_2 +\delta_{12}& l_1+\delta_{13} \\
			l_1l_2 +\delta_{12}& l_2^2 & l_2+\delta_{23} \\
			l_1+\delta_{13} & l_2 +\delta_{23}& 1
		}
		\label{eq.MNdelta}
	\end{eqnarray}
	where the real-valued perturbations $\delta_{ij}$ (for $i< j$) encode the hierarchy corrections. For clarity, flavor superscript $^q$ on $l_i$ and $\delta_{ij}$ is omitted when unambiguous.
	
\begin{figure}[H] 
		\centering  
		\vspace{-0.35cm} 
		\subfigtopskip=2pt 
		\subfigbottomskip=2pt 
		\subfigcapskip=-5pt 
		\subfigure[$l_1=1,l_2=1$]{
			\label{fig.so2masseigenvaluessub1}
			\includegraphics[width=0.25\linewidth]{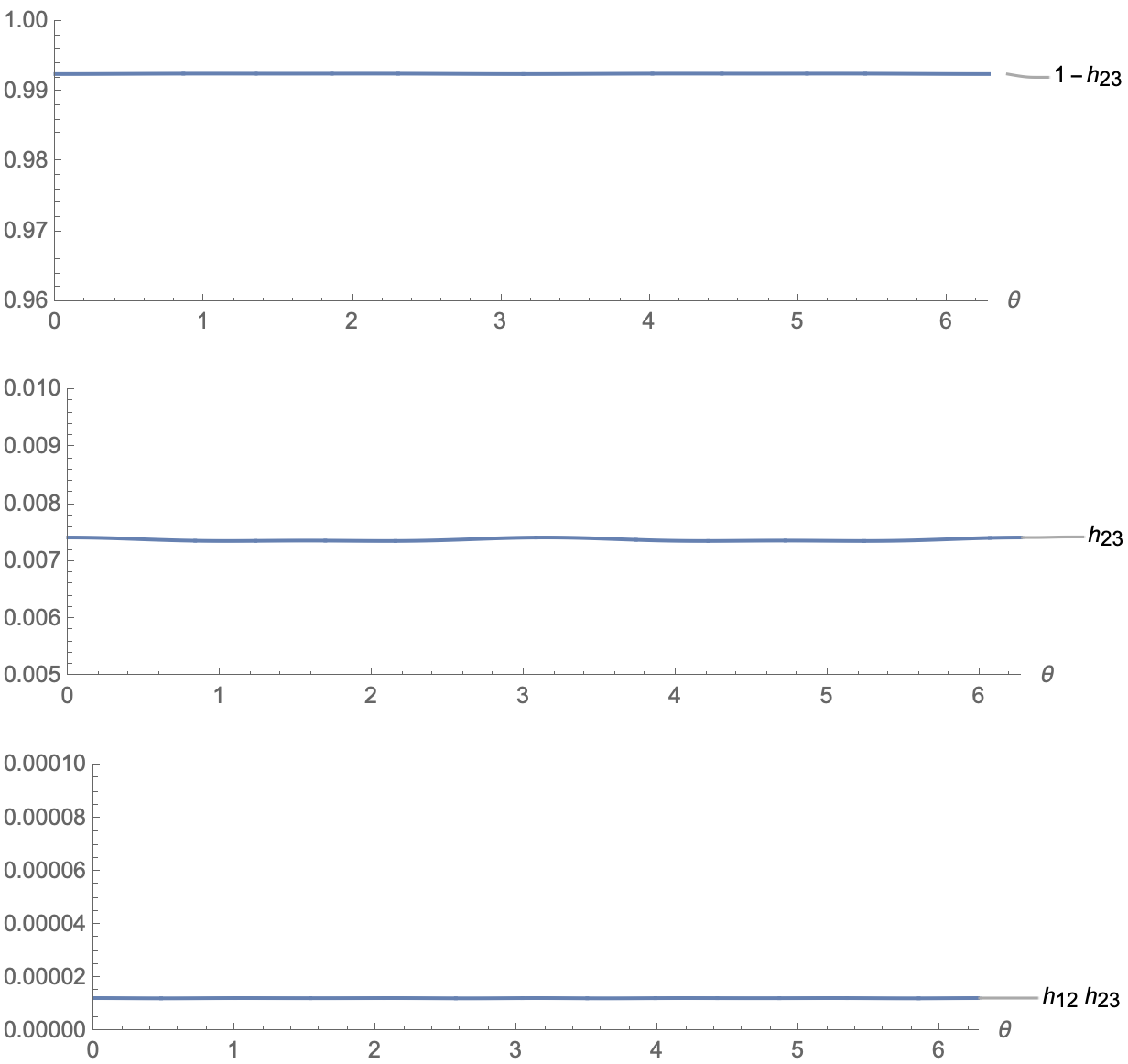}}
		\quad 
		\subfigure[$l_1=0.5,l_2=0.5$]{
			\label{fig.so2masseigenvaluessub2}
			\includegraphics[width=0.25\linewidth]{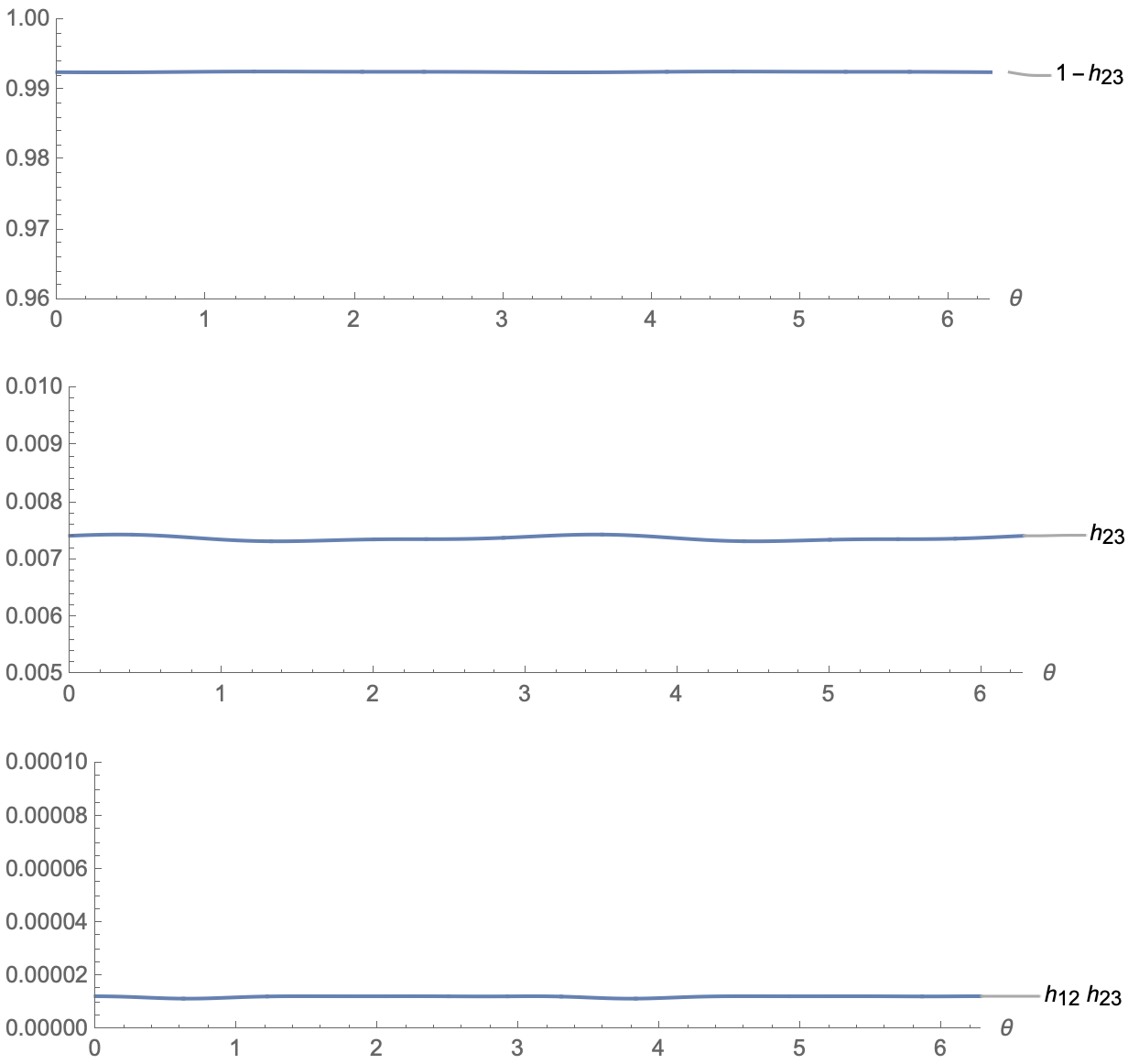}}
		\subfigure[$l_1=0.3,l_2=0.5$]{
			\label{fig.so2masseigenvaluessub3}
			\includegraphics[width=0.25\linewidth]{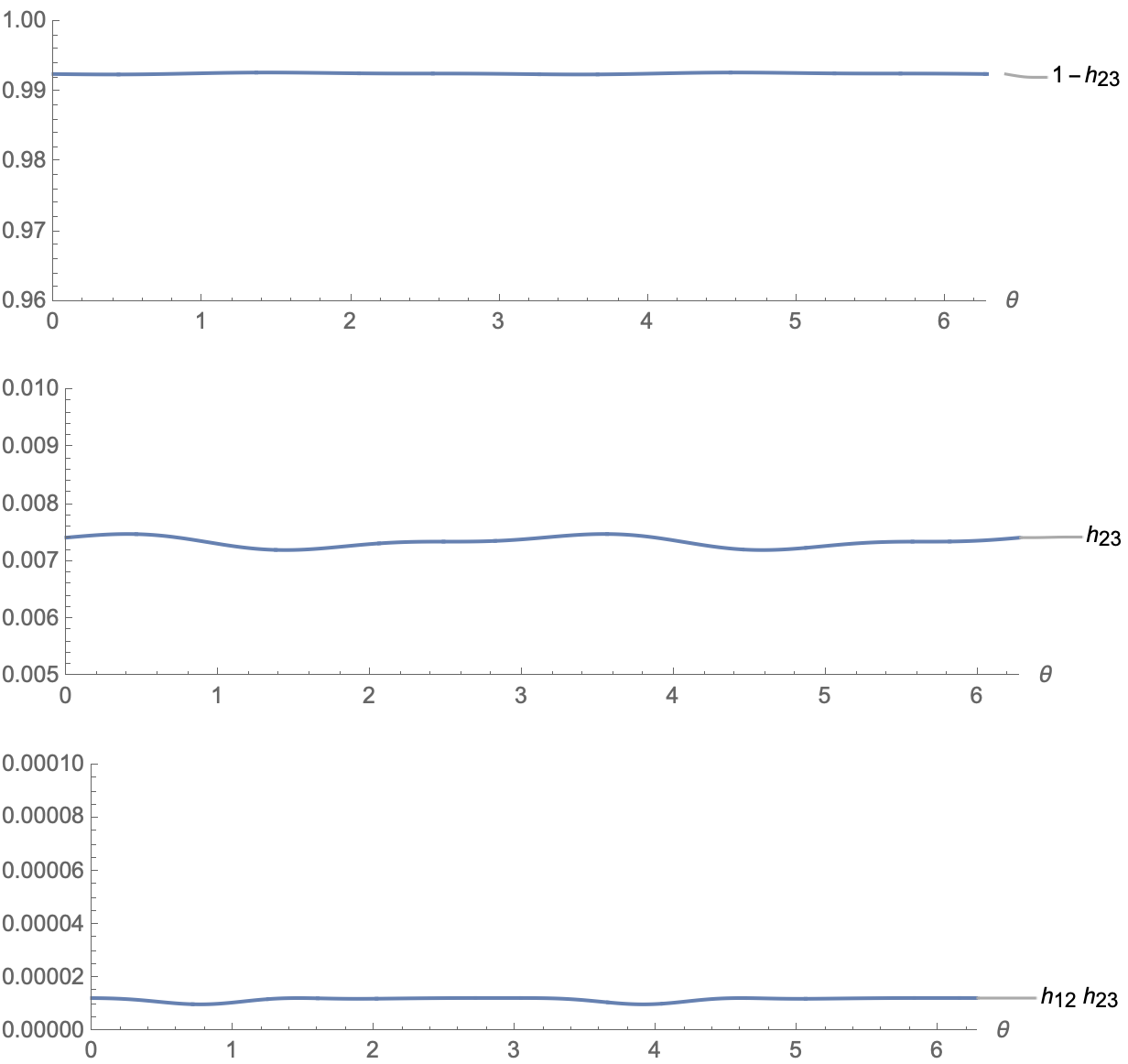}}
		\caption{Mass Eigenvalues of $M_N^q$ vs. rotation angle $\theta$.}
		\label{fig.so2masseigenvalues}
	\end{figure}

	To ensure the mass matrix produces the desired hierarchical eigenvalues $(h_{12}^qh_{23}^q, h_{23}^q, 1-h_{23}^q)$ at leading order of each family, the perturbation terms $\delta_{ij}$ must obey the following relations
	\begin{eqnarray}
		\delta_{12}^q&=&\mathcal{C}_{30}h_{23}^q+\mathcal{C}_{31}h_{12}^qh_{23}^q+\mathcal{C}_{32}(h_{23}^q)^2+\mathcal{O}(h^3),
		\nonumber\\
		\delta_{23}^q&=&\mathcal{C}_{10}h_{23}^q+\mathcal{C}_{11}h_{12}^qh_{23}^q+\mathcal{C}_{12}(h_{23}^q)^2+\mathcal{O}(h^3),
		\nonumber\\
		\delta_{13}^q&=&\mathcal{C}_{20}h_{23}^q+\mathcal{C}_{21}h_{12}^qh_{23}^q+\mathcal{C}_{22}(h_{23}^q)^2+\mathcal{O}(h^3).
		\label{eq.deltaij}
	\end{eqnarray}
	Here, $\mathcal{O}(h)$ coefficients are
	\begin{eqnarray*}
		\mathcal{C}_{30}&=&-\frac{1}{4 l_1 (1 + l_1^2) l_2}\Big\{(1 + l_1^2 + l_2^2) \Big[(1 + l_1^2) (l_1^2 + l_2^2) + (l_1^4 - l_2^2 + 
		 l_1^2 + l_1^2l_2^2) \cos(2 \theta) + 2 l_1 l_2 \sqrt{1 + l_1^2 + l_2^2} \sin(2 \theta)\Big]\Big\}
		\\
		\mathcal{C}_{10}&=&\frac{1}{4 (1 + l_1^2) l_2}(1 + l_1^2 + l_2^2) \Big\{-(1 + l_1^2) (1 + l_2^2) + (-1 - l_2^2 -l_1^2 + l_1^2l_2^2) \cos(
		2\theta) + 2 l_1 l_2 \sqrt{1 + l_1^2 + l_2^2} \sin(2\theta)\Big\}
		\\
		\mathcal{C}_{20}&=&-\frac{1}{4l_1}(1 + l_1^2) (1 + l_1^2 + l_2^2) \Big[1 - \cos(2\theta)\Big]
	\end{eqnarray*}
	Higher order coefficients $\mathcal{C}_{i1}$ and $\mathcal{C}_{i2}$  are provided in Appendix \ref{app.deltah2}. Here, $\theta$ is a $SO(2)$ rotation angles.
	For up-type quarks ($h_{23}^u=1.273/173,h_{12}^u=2.16/1273$), Fig. \ref{fig.so2masseigenvalues}  shows the $\theta$-dependence of $M_N^u$ eigenvalues in for the different $(l_1,l_2)$. The eigenvalues remain invariant under $SO(2)^q$ rotation up to $\mathcal{O}(h^2)$. It means that the corrections preserve an approximate $SO(2)^q$ family symmetry. 

	Now, let us count all parameters in the factorized structure with hierarchy corrections. There exist two diagonal phases $\eta_i$ in $K_L^q$ and three perturbations $\delta_{ij}^q$ in normalized $M_N^q$. Two of three perturbations $\delta_{ij}^q$ are necessary to  determine two mass hierarchies $h_{23}^q,h_{12}^q$, while leaving one free rotation angle $\theta$. 
	The total quark mass $\sum_i m_i^q$ requires a family universal Yukawa coupling $y^q$ in the mass term, expressed as 
	$$y^q\frac{v_0}{\sqrt{2}}=\sum m_i^q.$$
	Notably, the parameters $l_1$ and $l_2$, which are independent of quark masses, govern the mass pattern. 
	Upon choosing a specific pattern, $l_1$ and $l_2$ become fixed quantities.
	Consequently, the total parameter count for both up- and down-type quarks amounts to 12 (comprising 4 $\eta_i^{u,d}$,  6 $\delta_{ij}^{u,d}$, and 2 $y^{u,d}$). 
	Phenomenologically, these flavor parameters must reproduce 6 quark masses, 3 mixing angles, and 1 CP-violating phase, totalling 10 observables. This reveals that two of the twelve parameters in the factorized quark mass matrix $M^u$ and $M^d$ are redundant.
	As will be demonstrated in Sec. \ref{sec.mixing}, it will be proved that only two phase differences $\lambda_1=\eta_1^u-\eta_1^d$ and $\lambda_2=\eta_2^u-\eta_2^d$ are physically significant for the CKM mixing. Thus, Eq. (\ref{eq.KMK2}) establishes a faithful, non-redundant structure for describing flavor phenomenology. 
	The remaining task is to verify the effectiveness of this factorized structure by CKM mixing fits in Sec. \ref{sec.mixing}.

	\section{Patterns}
	\label{sec.AllPatterns}
	Historically, various patterns of mass matrices have been proposed in the study of flavor structures. Some originate from approximate symmetries of mixing parameters, others emerge from mass generation mechanisms \cite{WeinbergPRD2020}, while more are motivated by attempts to extend the SM through large symmetry groups \cite{JNLuPRD2018,AECHEPJC2016}. 
	Among these, texture zeros have gained particular popularity as they introduce specific vanishing elements in the mass matrix. The primary motivations for considering texture zeros typically involve simplifying flavor structures and/or incorporating symmetry principles.		
	The factorized mass structure presented in Eq. (\ref{eq.KMK2}) is fundamentally based on the general mass hierarchy. This approach has universal applicability and can encompass any mass patterns proposed to explain the characteristics of the mass hierarchy. Below, we examine some prominent patterns from the perspective of our factorized framework.

	\subsection{Fritzsch Pattern}
	First proposed by Fritzsch in 1979  to relate quark masses and the CKM mixing, this pattern assumes the following quark mass matrix \cite{FritzschNPB1979}
	\begin{equation}
		M^q = \begin{pmatrix}
			0 & a & 0 \\ 
			a & 0 & b \\ 
			0 & b & c
		\end{pmatrix}.
		\label{eq.FritzschM}
	\end{equation}
	While requiring modifications to accommodate current precision data, this pattern predicted relationships between quark masses and the CKM matrix elements in its original formulation. The matrix extends the seesaw mechanism to a sequential form: second family acquires its mass through the off-diagonal correction $b$, which then propagates to the first family by the correction $a$. When normalized to the total mass
	\begin{equation}
		M^q \sim\begin{pmatrix}
			0 & \frac{a}{c} & 0 \\ 
			\frac{a}{c} & 0 & \frac{b}{c} \\ 
			0 & \frac{b}{c} & 1
		\end{pmatrix},
		\label{eq.FritzschMn}
	\end{equation}
	it corresponds to our factorized structure with $l_1=l_2=0$. Here, the off-diagonal $a/c$ and $b/c$ represent $\delta_{12}^q$ and $\delta_{23}^q$ of Eq. (\ref{eq.MNdelta}) respectively.
	The Fritzsch pattern sets $\delta_{13}^q=0$, which fails to properly account for the $SO(2)$ symmetry crucial for flavor mixing (more details are discussed in the next section).
	This deficiency can be revealed by counting the number of parameters.
	After accounting for normalization, $M^q$ is expressed in terms of two parameters 
	$a/c$ and $b/c$. 
	These two parameters determine both mass hierarchies 
	$h_{12}^q$ and $h_{23}^q$, leaving no free parameters for CKM mixing unless 
	$a$ and $b$ are complex. However, complexifying these parameters violates the rule established in Eq. (\ref{Eq.rules1}).

	\subsection{Democratic Pattern}
	The democratic matrix represents a special case where all elements are nearly equal, historically employed to explain the significant mass hierarchies observed in quarks and leptons. The unperturbed matrix takes the form:
	\begin{eqnarray}
		M^q\sim \frac{1}{3}\Array{ccc}{1 & 1 &1 \\ 1 & 1 &1 \\ 1 & 1 &1 }
	\end{eqnarray}
	This corresponds to the $l_1=l_2=1$ case. 
	Diagonalization yields one massive fermion and two massless states, necessitating the introduction of small symmetry-breaking terms to generate masses for the lighter generations. A modification is often considered by introducing diagonal perturbations $\epsilon_1,\epsilon_2$ \cite{FritzschPPNG1999}
	\begin{eqnarray}
		M^q\sim \frac{1}{3}\Array{ccc}{1 & 1 &1 \\ 1 & 1 &1 \\ 1 & 1 &1 }+\frac{1}{3}\Array{ccc}{\epsilon_1 & 0 & 0 \\ 0 & \epsilon_2 & 0 \\ 0 & 0 & 0 }
	\end{eqnarray}
	By redefining $l_1=\sqrt{1+\epsilon_1}$ and $l_2=\sqrt{1+\epsilon_2}$, these diagonal pertrubations can be transformed into non-diagonal terms 
	\begin{eqnarray}
		M^q\sim \frac{1}{3}\Array{ccc}{l_1^2 & l_1l_2 & l_1 \\ l_1l_2 & l_2^2 & l_2 \\ l_1 & l_2 &1 }
		+\frac{1}{3}\Array{ccc}{0 & -\frac{\epsilon_1+\epsilon_2}{2} & -\frac{\epsilon_1}{2}
			\\ 
			-\frac{\epsilon_1+\epsilon_2}{2} & 0 & -\frac{\epsilon_2}{2} 
			\\  
			-\frac{\epsilon_1}{2} & -\frac{\epsilon_2}{2}  & 0 }+\mathcal{O}(\epsilon^2).
	\end{eqnarray}
	It establishes relationships between perturbations in the general factorized pattern and the democratic case
	\begin{eqnarray}
		\delta_{12}&=&-\frac{\epsilon_1+\epsilon_2}{2},
		\\
		\delta_{13}&=&-\frac{\epsilon_1}{2},
		\\
		\delta_{23}&=&-\frac{\epsilon_2}{2}.
	\end{eqnarray}
	Like the Fritzsch texture, the democratic matrix suffers from insufficient handling of $SO(2)$ symmetry due to having only two free parameters to describe both mass hierarchies. 
	By directly complexifying $\epsilon_i$, the democratic pattern might address CP violation. 
	However, it also would violate the rules in Eq. (\ref{Eq.rules1}). Given these differences, we sometimes denote the case with $l_1=l_2=1$ as the "flat pattern" to distinguish it from traditional democratic matrices in the manner of perturbation corrections.

	\subsection{Triangular Texture}
	Triangular texture with 6 zero elements is used to explain hierarchy by sequential symmetry breaking
	\begin{eqnarray}
		M_q = \begin{pmatrix}
			0 & a & 0 \\ 
			0 & 0 & b \\ 
			0 & 0 & c
		\end{pmatrix}
	\end{eqnarray}
	Non-physical right-handed transformations need to be eliminated by considering $M_qM_q^\dag$. Normalizing to the total mass, $M_qM_q^\dag$ becomes
	\begin{eqnarray*}
		M_qM_q^\dag
		\sim \Array{ccc}{(\frac{a}{c})^2 & 0 & 0 \\ 0 & (\frac{b}{c})^2 & \frac{b}{c} \\ 0 & \frac{b}{c} & 1}
	\end{eqnarray*}
	Since the condition $l_1l_2=0$ is imposed at the matrix position $(2,1)$, the diagonal term $(a/c)^2$, which violates the consistency relation in Eq. (\ref{Eq.rules2}), must necessarily originate from perturbation effects. It can be realized equivalently by introducing a non-diagonal complex $\delta_{12}$. Thus, the triangular texture is incorporated into $M_N^q$ with $ l_1=0$ and $ l_2=b/c$.

	\subsection{Patterns from the Mixing}
	Various flavor patterns have been proposed based on mixing phenomenology, particularly those motivated by observed structures in the PMNS matrix, including the $Z_2$-symmetric texture arising from $\mu$-$\tau$ symmetry. However, as these patterns primarily focus on mixing characteristics rather than explaining mass hierarchies, they are fundamentally incompatible with our factorized mass structure.
	
	A prominent example is the tribimaximal mixing (TBM) pattern, which was motivated by an observed approximate symmetry in neutrino mixing
	\begin{eqnarray}
		U_{TBM}=\Array{ccc}{\sqrt{\frac{2}{3}} & \sqrt{\frac{1}{3}} & 0  \\ -\sqrt{\frac{1}{6}} & \sqrt{\frac{1}{3}} & \sqrt{\frac{1}{2}}  \\ \sqrt{\frac{1}{6}}  & -\sqrt{\frac{1}{3}}  & \sqrt{\frac{1}{2}} }
	\end{eqnarray}
	corresponding to three  mixing angles: solar angle $\theta_{12}=\arcsin \sqrt{\frac{1}{3}}\simeq 35.3^\circ$, 
	atmospheric angle $\theta_{23}=45^\circ$, and reactor angle $\theta_{13}=0$.
	While originally formulated for Majorana neutrinos, this pattern can also be adopted for neutrinos \cite{HarrisonPLB2002, ZHZhaoJHEP2014}. 
	The TBM pattern typically emerges from discrete flavor symmetry breaking, where neutrinos and charged leptons transform under specific group representations (e.g., $A_4$, $S_4$) \cite{KingJHEP2011}. 
	The simplest neutrino mass matrix consistent with TBM takes the form
	\begin{eqnarray*}
		M_\nu=\frac{m_2}{3}\Array{ccc}{1& 1 &1 \\ 1& 1 &1 \\ 1& 1 &1 }+\frac{m_3}{2} \Array{ccc}{0 & 0 & 0 \\ 0 & 1 & -1 \\ 0 & -1 & 1}.
	\end{eqnarray*}
	This PMNS-inspired pattern can only be accommodated within our $M_N^q$ framework under an additional condition $m_2 \ll m_3$.

	\section{Flavor Mixing}
	\label{sec.mixing}
	
	While the mass matrix in Eq. (\ref{eq.KMK2}) provides a fundamental structure, the flavor mixing matrix cannot be uniquely determined without resolving two key conceptual challenges: (1) the selection of an appropriate quark mass basis, and (2) the mechanism of family symmetry breaking induced by charged-current weak interactions. These issues require careful treatment, as past analyses have prematurely excluded otherwise viable mixing patterns due to insufficient consideration of non-zero $\theta_{13}$, CP-violating effects, or the large $\theta_{12}$  observed in PMNS mixing.
	
	A rigorous approach to family symmetry breaking is particularly crucial, as it can substantially modify the predicted mixing patterns and restore compatibility with experimental observations. The historical misjudgment of certain patterns underscores the importance of properly accounting for these symmetry breaking effects, which may otherwise obscure potentially valid solutions to the flavor puzzle. 
	
	\subsection{The Choice of Basis}	
	The quark mixing matrix in the charged weak current, $U_{CKM}=U_L^u(U_L^d)^\dag$, emerges from the mismatch between the left-handed rotations of up- and down-type quarks. Some studies simplify this structure by assuming up-type quarks as common eigenstates of both weak gauge interactions and masses, thereby reducing the CKM matrix to $U_{CKM}=(U_L^d)^\dag$. However, this simplification artificially decouples the up-type quark masses from their contribution to flavor mixing and masks the underlying mechanisms that generate up-type quark mass hierarchy.
	A robust flavor structure must consistently explain all quark masses and mixing, without resorting to biased basis choices that obscure part of the physical picture.
	
	When constructing the mass matrices $M_N^u$ and $M_N^d$, two distinct strategies arise:
	\begin{itemize}	
		\item[(1)] sector specific pattern parameters. This method assigns distinct texture parameters (e.g., $l_1,l_2$) to up- and down-type quarks, permitting independent patterns in each sector. While this flexibility can accommodate observed mass hierarchies and mixing, it introduces additional degrees of freedom without necessarily explaining their origin.
		\item[(2)] universal pattern parameters. Alternatively, one may adopt a unified set of parameters (e.g., common $l_1,l_2$) for both quark types, suggesting a shared dynamical origin for their flavor structure. This approach is more predictive and theoretically appealing, as it reduces arbitrariness and aligns with the notion that a fundamental theory should unify quark mass family.
	\end{itemize}
	We advocate for the second strategy, as it not only minimizes free parameters but also provides a more coherent framework for understanding quark masses and mixing within a single theoretical paradigm. Such universality could point to deeper symmetries or mechanisms governing flavor physics.

	\subsection{$SO(2)$ Family Symmetry Breaking}
	Another crucial aspect of flavor structure involves understanding how family symmetry breaking manifest in quark mixing. 
	A common misconception suggests that in the limit of quark mass degeneracy, the CP-violating phase could be removed through rotations in the degenerate subspace \cite{JarlskogPRL1985}. We address this by first examining the hierarchy limit before considering corrections.
	
	In the hierarchy limit, the first two families possess an $SO(2)^q$ symmetry, This permits arbitrary rotations $R_3(\theta)$ within the degenerate subspace
	\begin{eqnarray*}
		R_3(\theta)=\Array{ccc}{c_\theta & -s_\theta & 0 \\ s_\theta & c_\theta & 0 \\ 0 & 0 & 1}.
	\end{eqnarray*}
	with $s_\theta=\sin\theta$ and $c_\theta=\cos\theta$.
	Notably, if $U_L^q$ diagonalizes $M^q$
	\begin{eqnarray}
		U_L^qM^q(U_L^q)^\dag=\textrm{diag}(m_1^q,m_2^q,m_3^q),
	\end{eqnarray}
	$R_3(\theta)U_L^q$ equally serves as a valid diagonalization matrix
	\begin{eqnarray}
		R_3(\theta)U_L^qM^q\Big[R_3(\theta)U_L^q\Big]^\dag=\textrm{diag}(m_1^q,m_2^q,m_3^q).
	\end{eqnarray}
	Consequently, the CKM matrix acquires additional rotational freedom
	\begin{eqnarray}
		U_{CKM}=R_3(\theta^u)U_L^u[U_L^d]^\dag [R_3(\theta^d)]^\dag
	\end{eqnarray}
	Critically, quark rephasing cannot eliminate these rotational effects, revealing that the charged weak current explicitly breaks the $SO(2)^q$   family symmetry.	
	
	Two key phenomenological observations support the breaking of $SO(2)^{u,d}$ symmetry. 
	First, the structure of CKM mixing itself provides evidence: if the $SO(2)^q$ rotation angles were truly free parameters, the CKM matrix would exhibit arbitrary mixing angles and CP-violating phases constrained only by $SO(2)^{u,d}$ symmetry. For instance, the Cabibbo angle $\theta_{12}$ would be unrestricted in semileptonic kaon decays and charmless B decays, contrary to experimental measurements that show it taking specific, well-determined values.
	
	Second, the observed CP violation in quark sector presents a compelling case. In the hierarchy limit where $h_{23}^q \rightarrow 0$, the CP-violating phase $\delta_{CP}$ can be expressed through the expansion
	$$\delta_{CP}=\delta_{CP,0}+c_1h_{23}^1+c_2h_{23}^2+c_3h_{23}h_{12},$$
	where all hierarchy-dependent terms vanish, leaving only $\delta_{CP,0}$. 
	This suggests that while hierarchy corrections contribute to CP violation, the measured value of $\delta_{CP} \approx 61^\circ$ cannot be explained solely by these corrections when the hierarchy limit remains a good approximation. The significant deviation from zero requires that the $SO(2)^{u,d}$ mixing angles $\theta^u$ and $\theta^d$ be fixed quantities rather than free parameters, as confirmed by experimental data on CKM mixing.

	\subsection{The CKM mixing}
	The unperturbed mass matrix $M_N^q$ is diagonalized by  transformation  $S_0$
	\begin{eqnarray}
		S_0M_N^q(S_0^q)^T=\textrm{diag}(0,0,1)
	\end{eqnarray}
	with explicit form
	\begin{eqnarray*}
		S_0= 
		\begin{pmatrix}
			\frac{1}{\sqrt{1+l_1^2}} & 0 & -\frac{l_1}{\sqrt{1+l_1^2}} \\
			-\frac{l_1l_2}{\sqrt{(1+l_1^2)(1+l_1^2+l_2^2)}} & \frac{\sqrt{1+l_1^2}}{\sqrt{1+l_1^2+l_2^2}} & -\frac{l_2}{\sqrt{(1+l_1^2)(1+l_1^2+l_2^2)}} \\
			\frac{l_1}{\sqrt{1+l_1^2+l_2^2}} & \frac{l_2}{\sqrt{1+l_1^2+l_2^2}} & \frac{1}{\sqrt{1+l_1^2+l_2^2}} 
		\end{pmatrix}.	
	\end{eqnarray*}
		The general diagonalization includes $SO(2)^q$ rotation $R_3(\theta)$
	\begin{equation}
		\left[R_3(\theta)S_0\right] M_N \left[R_3(\theta)S_0\right]^\dag = \text{diag}(0,0,1).
		\label{eq.RMR10}
	\end{equation}
	Using Eq. (\ref{eq.KMK2}) and Eq. (\ref{eq.RMR10}), the left-handed quark transformation becomes
	\begin{equation}
		U_L^q = R_3(\theta^q) S_0 K_L^q,
		\label{eq.Ulq01}
	\end{equation}
	yielding the CKM mixing matrix
	\begin{equation}
		U_{CKM} = R_3(\theta^u) S_0 ~\textrm{diag}(e^{i\lambda_1}, e^{i\lambda_2}, 1) S_0^T R_3^T(\theta^d),
		\label{eq.Uckm0}
	\end{equation}
	The phase structure of the CKM matrix reverls an elegant reduction of parameters. The original four phases $\eta_{1,2}^{u,d}$ in $K_L^{u,d}$ comine into just two physical phase differences
	\begin{eqnarray}
		\lambda_i \equiv \eta_i^u - \eta_i^d.
	\end{eqnarray}
	This occurs because relative phases between up- and down-type quarks are observable, while overall sector phases cancel in the CKM construction.
	
	The $SO(2)^q$ rotation angles $\theta^q$ appear as free parameters in $M_N^q$ diagonalization.
	However, they become physically meaningful when constrained by quark mixing in weak interactions. 
	Eq. (\ref{eq.Uckm0}) demonstrates that the CKM matrix is completely determined by just four parameters: $\theta^u$, $\theta^d$, $\lambda_1$, and $\lambda_2$. This minimal parameterization elegantly matches the experimental reality of three mixing angles and one CP-violating phase in quark mixing.

	The complete diagonalization transformation $R_D^q$ for the hierarchy corrected $M_N^q$ in Eq. (\ref{eq.MNdelta}) satisfies
	\begin{eqnarray}
		R_D^qM_N^q(R_D^q)^T=\textrm{diag}(h_{12}^qh_{23}^q,h_{23}^q,1-h_{23}^q)+\textrm{high order terms}.
	\end{eqnarray}
	As confirming consistensy with the $SO(2)^q$ symmetry limit, the transformation $R_D^q(\theta)$ properly reduces to the unperturbed case
	\begin{eqnarray}
		R_D^q(\theta)\xrightarrow{h_{23}^q\rightarrow0}R_3(\theta)S_0.
	\end{eqnarray}
	Further analysis reveals that the $SO(2)^q$ family symmetry remains a good approximate symmetry even when hierarchy corrections are included. The complete transformation $R^q(\theta)$ can be systematically decomposed into hierarchy-corrected components
	\begin{eqnarray}
		R_D^q(\theta)&=&\tilde{R}_3(\theta)	\tilde{S}
		\\
		\tilde{R}_3(\theta)&=&R_3(\theta)+R'_{1}(\theta)h_{23}^q+R'_{2}(\theta)h_{23}^qh_{12}^q+R'_{2}(\theta)(h_{23}^q)^2+\cdots
		\\
		\tilde{S}&=&S_0+S_{1}h_{23}^q+S_{1}h_{23}^qh_{12}^q+S_{2}(h_{23}^q)^2+\cdots
	\end{eqnarray} 	 
	For the specific case of the flat pattern where $l_1=l_2=1$, the explicit forms of the correction coefficients $R'{i}$ and $S_i$ are provided in reference \cite{Zhang2025arxiv}. Remarkably, despite these hierarchy corrections, the fundamental structure of the CKM mixing matrix remains unchanged
	\begin{eqnarray}
		U_{CKM}=R_D(\theta^u)\text{diag}(e^{i\lambda_1},e^{i\lambda_2},1)R_D^T(\theta^d).
		\label{eq.UckmDelta}
	\end{eqnarray}

	\section{Flavor Mixing Fit}
	\label{sec.fit}
	We now focus on fitting the two $SO(2)^f$ rotation angles and CP-violating phases $\lambda_1,\lambda_2$  to CKM and PMNS mixing data.

	\subsection{The CKM Mixing Fit}
	The CKM mixing is parameterized in a standard form with three mixing angles and one CP-violating phase 
	\begin{eqnarray}
		U_{CKM}=\begin{pmatrix}1 & 0 & 0 \\ 0 & c_{23} & s_{23} \\ 0 & -s_{23} & c_{23}\end{pmatrix}
		\begin{pmatrix}c_{13} & 0 & s_{13}e^{-i\delta_{CP}} \\ 0 & 1 & 0 \\ -s_{13}e^{i\delta_{CP}} & 0 & c_{13}\end{pmatrix}
		\begin{pmatrix}c_{12} & s_{12} & 0 \\ -s_{12} & c_{12} & 0 \\ 0 & 0 & 1\end{pmatrix}
	\end{eqnarray}
	where $s_{ij}\equiv \sin\theta_{ij}$ and $c_{ij}\equiv \cos\theta_{ij}$.
	The three mixing angles $\theta_{ij}$ can be directly extracted from the unphased CKM matrix elements by the relations
	\begin{eqnarray}
		s_{13}&=&|U_{CKM,13}|,
		\label{eq.s13Cal}
		\\
		s_{23}^2&=&\frac{|U_{CKM,23}|^2}{1-|U_{CKM,13}|^2},
		\label{eq.s23Cal}
		\\
		s_{12}^2&=&\frac{|U_{CKM,12}|^2}{1-|U_{CKM,13}|^2}.
		\label{eq.s12Cal}
	\end{eqnarray}
	The CP-violating phase $\delta_{CP}$ is determined via Jarlskog invariant \cite{JarlskogPRL1985}
	\begin{eqnarray}
		J_{CP}&=&\textrm{Im}[U_{CKM,11}U_{CKM,22}U_{CKM,12}^*U_{CKM,21}^*]
		\nonumber\\
		&=&s_{13}c_{13}^2s_{23}c_{23}s_{12}c_{12} \sin\delta_{CP}.
		\label{eq.Jcpmixingangle}
	\end{eqnarray}
	
	Before performing the CKM mixing fit, we must first specify the mass matrix pattern by fixing the parameters $l_1$ and $l_2$. The quark mass hierarchies $h_{ij}^{u,d}$ are taken from experimental mass data. The CKM matrix then depends on four remaining parameters: $\theta^u$, $\theta^d$, $\lambda_1$, and $\lambda_2$.
	For any given parameter combination ($\theta^u$, $\theta^d$, $\lambda_1$, $\lambda_2$), we compute the corresponding mixing angles and CP-violating phase using Eqs. (\ref{eq.s13Cal})-(\ref{eq.Jcpmixingangle}). The fit goodness is evaluated through the $\chi^2$ statistic 
	\begin{eqnarray}
		\chi^2=\sum_{O_i}\left(\frac{O_{i}-O_{i}^{(exp)}}{\delta O_{i}^{(exp)}}\right)^2
	\end{eqnarray}
	where $O_i$ represents the set of observables {$s_{12}$, $s_{23}$, $s_{13}$, $\delta_{\text{CP}}$}, with $O_i^{(\text{exp})}$ denoting their experimental values. 
	We retain all parameter points satisfying $\chi^2 < 4.6$, corresponding to $1\sigma$ C.L..
	
\begin{figure}[H] 
		\centering  
		\vspace{-0.35cm} 
		\subfigtopskip=2pt 
		\subfigbottomskip=2pt 
		\subfigcapskip=-5pt 
		\subfigure[$0\leq l_1, l_2,\leq 1$]{
			\label{fig.quark}
			\includegraphics[width=0.40\linewidth]{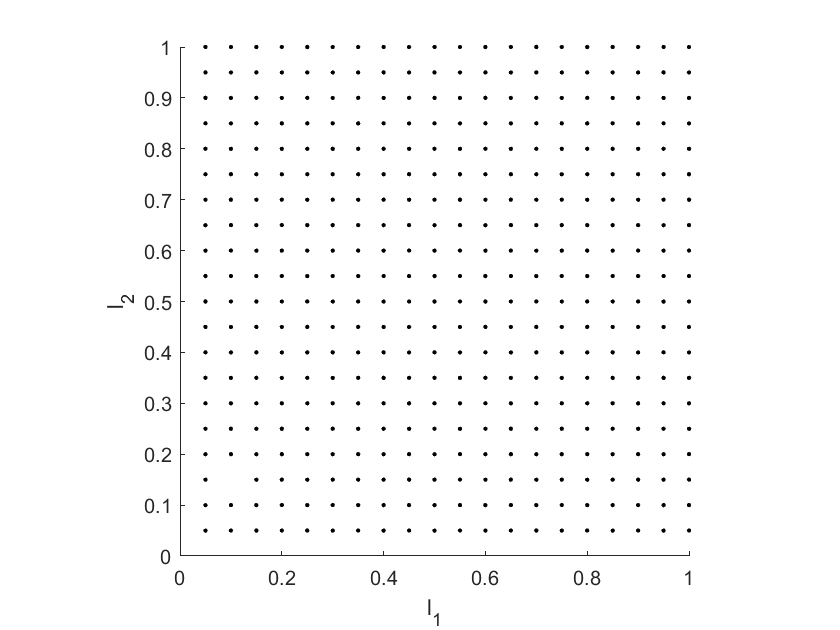}}
		\quad 
		\subfigure[$0\leq l_1\leq 1$ and $0\leq l2 \leq 0.05$]{
			\label{fig.quarksub1}
			\includegraphics[width=0.15\linewidth]{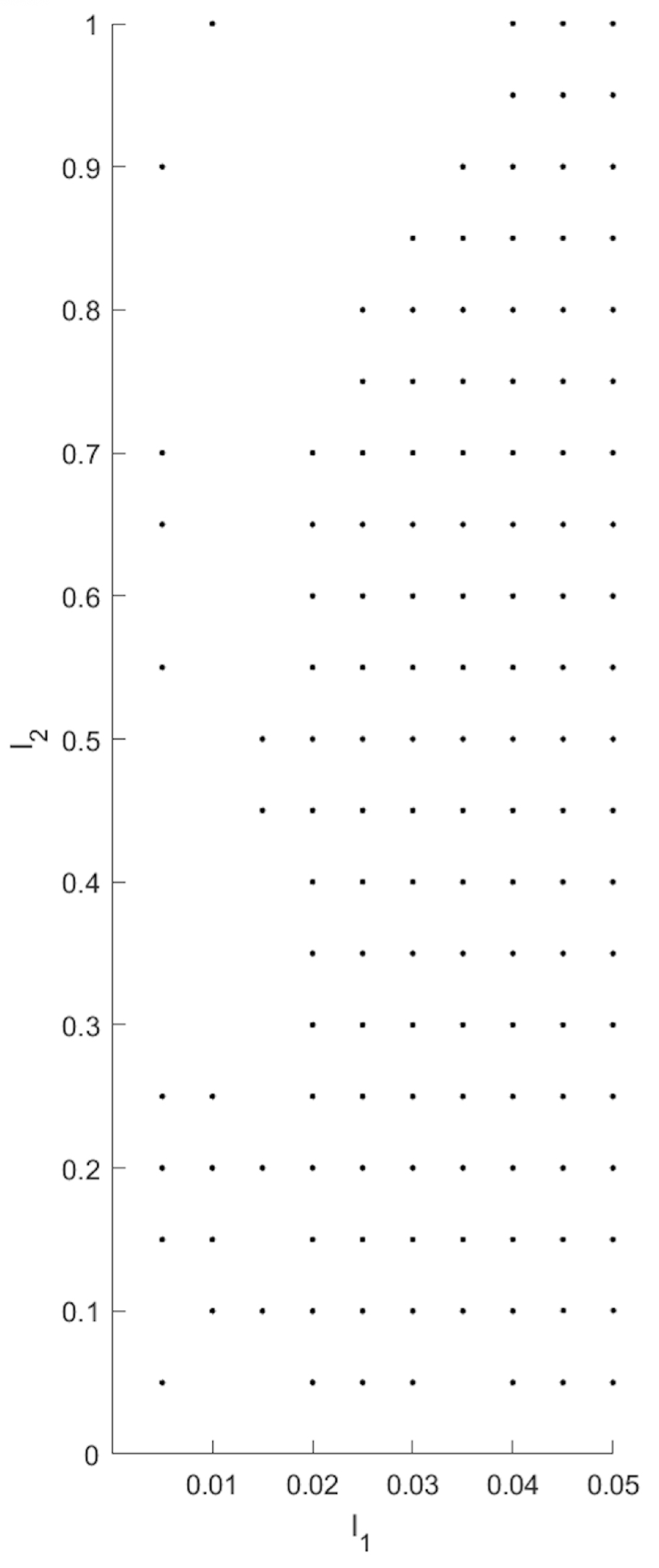}}
		\quad 
		\subfigure[$0\leq l_1\leq 0.05$ and $0\leq l_2 \leq 1$]{
			\label{fig.quarksub2}
			\includegraphics[width=0.30\linewidth]{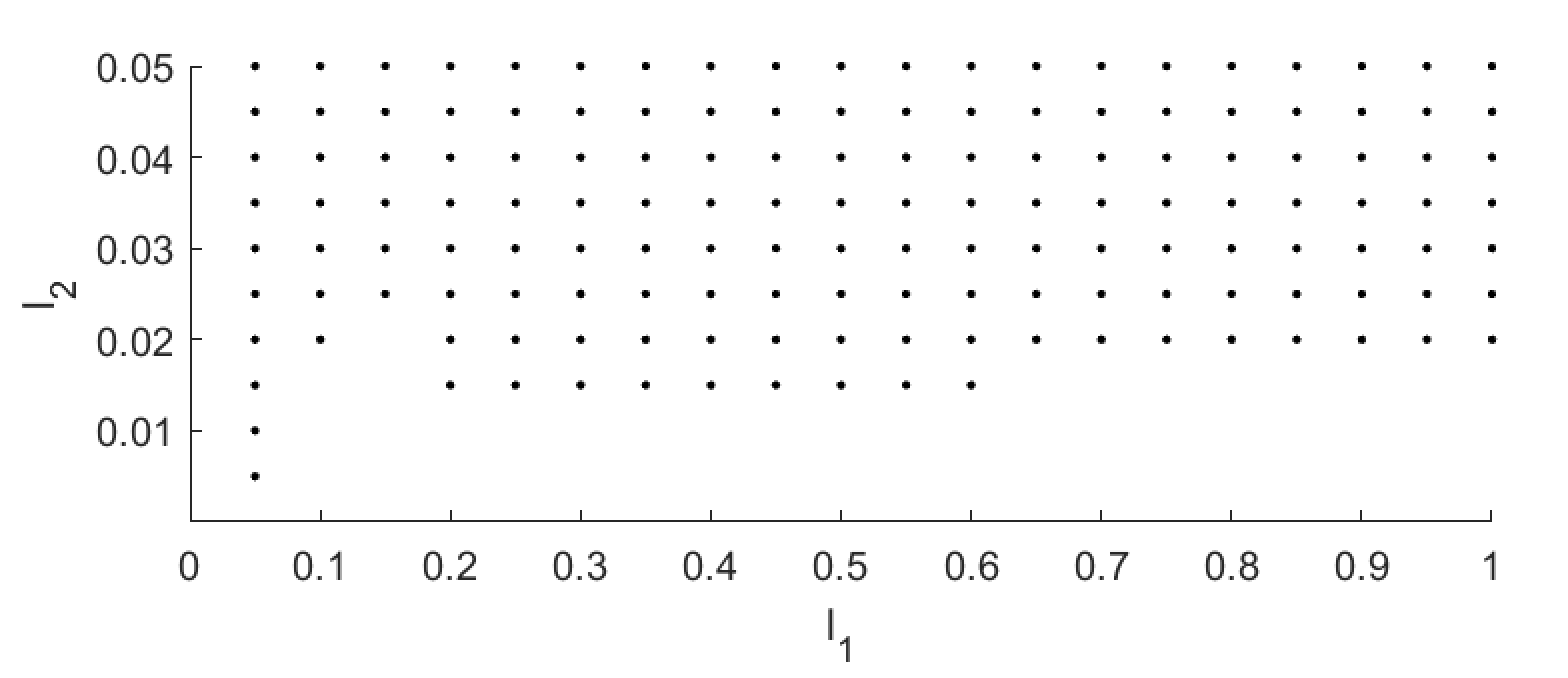}}
		\caption{Allowed quark mass patterns by the CKM mixing at $1\sigma$ in the plane of $(l_1,l_2)$. The quark hierarchies are taken $h_{23}^u=0.007370,h_{12}^u=0.001697, h_{23}^d=0.02273,h_{12}^d=0.05027$ from  2024 PDG data \cite{PDG2024}.}
		\label{fig.CKMfitPattern}
	\end{figure}

	Through a comprehensive scan of the ($l_1$,$l_2$) parameter space ($0 \leq l_1,l_2\leq 1$ with step length $0.05$), we identify all viable patterns that successfully reproduce the observed CKM mixing. The results are presented in Fig. \ref{fig.quark}, which maps the allowed patterns in the ($l_1$,$l_2$) plane.
	
	It is an important phenomenological observation that the current CKM mixing data allows nearly all patterns within our framework to provide acceptable fits at the $1\sigma$ C.L.. This flexibility originates from two sources: one is the relatively large experimental errors in certain mixing matrix elements, and another is the compensatory freedom introduced through the broken $SO(2)^q$ symmetry angles.
	
	Our fitting analysis yields a particularly significant exclusion to patterns with either $l_1=0$ or $l_2=0$, which are highlighted in Figs. \ref{fig.quarksub1} and \ref{fig.quarksub2}. 
	This robust exclusion implies that texture-zero patterns in $M_N^q$ are fundamentally incompatible with precision CKM mixing data.
	This result warrants particular attention because our treatment of both the quark mass basis transformation and $SO(2)^q$ symmetry breaking mechanism differs fundamentally from conventional approaches in the literature. 
	The more numerical results supporting these conclusions are provided in Appendix \ref{app.fitdata} to facilitate independent verification and further analysis.

	\subsection{The PMNS Mixing Fit}
	When extending the quark formalism to the lepton sector, several complications arise due to fundamental differences between these fermions. The primary challenge stems from the ambiguous nature of neutrino masses, particularly regarding their mass ordering. The factorized mass structure, which relies on mass hierarchy, is strictly applicable only to normal-ordered neutrinos. In the case of inverted ordering, the mass matrix reconstruction requires a distinct parameterization scheme that demands further investigation.
	Moreover, the PMNS mixing matrix for Majorana neutrinos differs significantly from the quark CKM mixing, as it incorporates two additional Majorana phases. Consequently, the structure of quark CKM mixing can only be directly translated to the lepton sector for Dirac neutrinos with normal mass ordering, achieved by substituting the quark labels with their lepton counterparts $u\rightarrow e$ and $d\rightarrow \nu$. 
	
\begin{figure}[H] 
		\centering  
		\vspace{-0.35cm} 
		\subfigtopskip=2pt 
		\subfigbottomskip=2pt 
		\subfigcapskip=-5pt 
		\subfigure[$h_{23}^{\nu}=0.173,h_{12}^{\nu}=0.114$]{
			\label{fit.PMNSfit01}
			\includegraphics[width=0.40\linewidth]{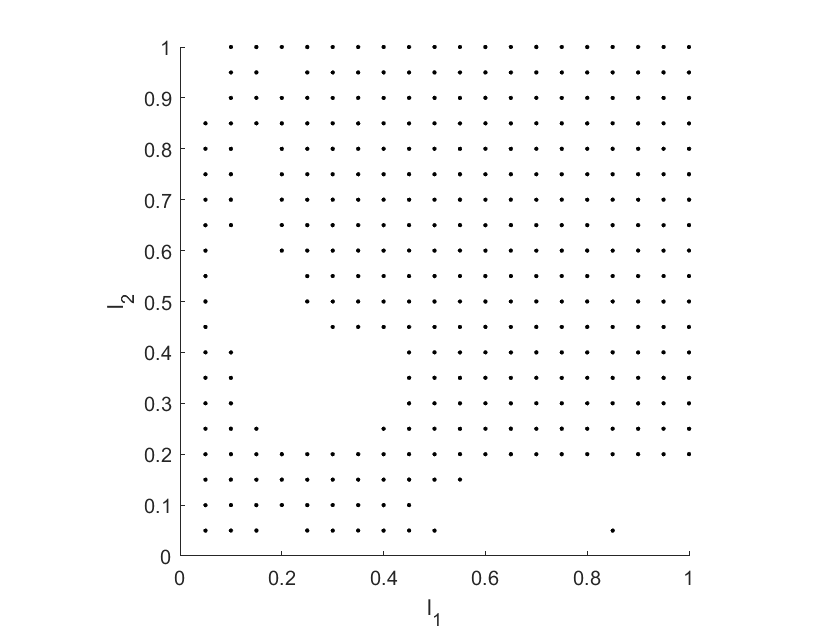}}
		\quad 
		\subfigure[$h_{23}^{\nu}=h_{12}^{\nu}=0.01$]{
			\label{fit.PMNSfit02}
			\includegraphics[width=0.40\linewidth]{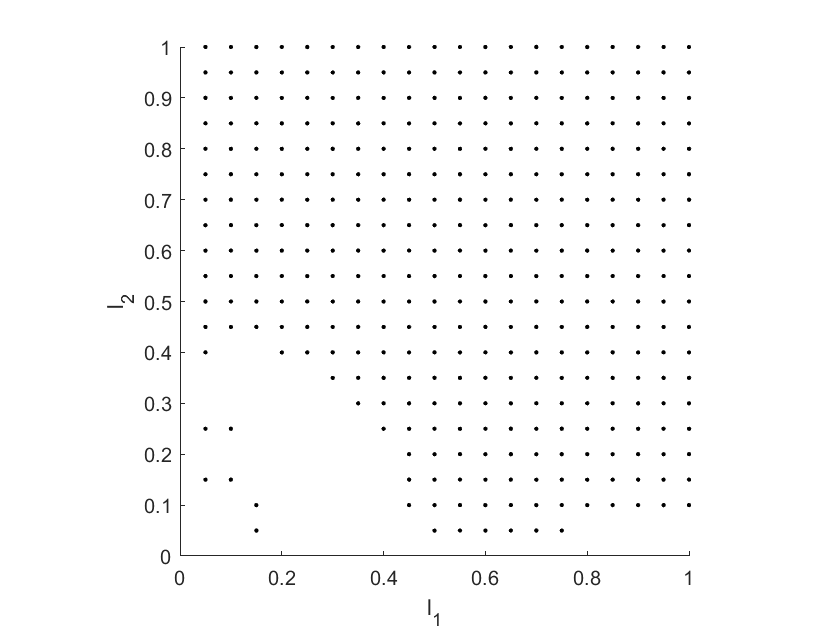}}
		\caption{All lepton mass patterns in $(l_1,l_2)$. The hierarchies are taken $h_{23}^e=0.007370,h_{12}^e=0.001697$ for charged leptons \cite{PDG2024}. }
		\label{fig.PMNSfitPattern}
	\end{figure}

	Regarding neutrino mass-squared differences, we adopt the hierarchy parameters $h_{23}^\nu=0.173,h_{12}^\nu=0.114$ based on the observed values of  $\Delta (m^\nu_{21})^2$ and $\Delta (m^\nu_{32})^2$. 
	These hierarchy parameters induce notable corrections to the PMNS mixing angles, which we compute up to $\mathcal{O}(h^2)$. The complete set of viable mixing patterns, incorporating these corrections, is presented in Fig. \ref{fit.PMNSfit01}.
	We also further examine an idealized scenario with minimal neutrino mass hierarchy by setting $h_{23}^{\nu}=0.01$ and $h_{12}^{\nu}=0.01$, the results of which are presented in Fig. \ref{fit.PMNSfit02}. In this regime, the contributions from neutrino mass hierarchy are strongly suppressed, leading to a simplified mixing pattern that closely resembles the leading-order approximation. This limiting case provides a useful reference for understanding the role of hierarchy effects in shaping the PMNS matrix.
	The more numerical results are also provided in Appendix \ref{app.fitdata}.
	
	\section{summary}
	Inspired by the distinctive mass hierarchies observed in quarks and charged leptons, we develop a factorized mass structure framework where the mass matrix decomposes as $K_L^fM_N^f$. This elegant separation isolates the flavor pattern, determined by the real symmetric matrix $M_N^f$, from CP-violating effects encoded in the diagonal phase matrix $K_L^f$. The framework's parameters precisely reproduce all essential physical observables, including mass hierarchies, mixing angles, and the CP-violating phase, while maintaining a minimal, non-redundant parameter form. Our analysis reveals that the flavor mixing mechanism originates from the breaking of an approximate $SO(2)^f$ family symmetry in the weak charged current. It is a conclusion supported by both theoretical considerations and robust phenomenological evidence. The resulting mixing matrix emerges as a clear composition of $SO(2)^f$ rotations and complex phase contributions, successfully describing both quark sector and lepton sector with extended normal-ordered Dirac neutrinos. Through comprehensive fitting to CKM and PMNS mixing data, we systematically explore the complete space of viable patterns parameterized by $(l_1,l_2)$. It demonstrates the framework's effectiveness, organization, and non-redundancy as a candidate for a fundamental flavor structure. While this approach provides significant explanatory power, some important questions remain regarding the dynamical origin of mass perturbations and their connection to Yukawa interactions, presenting compelling directions for future theoretical investigation.
	
	\section*{Acknowledgments}
	This work is supported by Shaanxi Foundation SAFS 22JSY035 of China.
	
	\begin{appendix}
		\section{The $h^2$-order Corrections of Mass Matrix}
		\label{app.deltah2}
		To address the first family quark mass, $h^2$-order corrections need to be calculated.
		The $\mathcal{O}(h^2)$ corrections $\delta_{ij}^q$ in Eq. (\ref{eq.MNdelta}) can be expressed as 
		\begin{eqnarray}
			\mathcal{C}_{31}&=&-\frac{1}{2l_1l_2}(l_1^2 + l_2^2) (1 + l_1^2 + l_2^2)
			\\
			\mathcal{C}_{32}&=&\frac{1}{(64 l_1 (1 + l_1^2)^2 l_2^3)} (1 + l_1^2 + l_2^2)^2\Big\{
			-(1 + l_1^2)^2 (l_2^2 + l_2^4 + l_1^2 (3 + l_2^2))
			\\
			&&-4 l_1^2 (1 + l_1^2) (1 + l_1^2 + 2 l_2^2) \cos(2\theta) 
			\\
			&&+\Big[l_2^2 + l_2^4 + l_1^6 (-1 + l_2^2) + l_1^4 (-2 - 5 l_2^2 + l_2^4) - l_1^2 (1 + 5 l_2^2 + 6 l_2^4)\Big] \cos(4\theta) 
			\\
			&&+4 l_1 (-1 + l1^2) l2 \sqrt{1+ l_1^2 + l_2^2} \Big[1 + l_1^2 + (1 + l_1^2 + 2 l_2^2) \cos(2\theta) \Big] \sin(2\theta) \Big\}
			\\
			\mathcal{C}_{11}&=&-\frac{1}{2l_2}(1 + l_1^2 + 2 l_2^2 + l_1^2 l_2^2 + l_2^4)
			\\
			\mathcal{C}_{12}&=&\frac{1}{l_1}\mathcal{C}_{32}
			\\
			\mathcal{C}_{21}&=&-\frac{1}{2l_1}(1 + l_1^2) (1 + l_1^2 + l_2^2)
			\\
			\mathcal{C}_{22}&=&0
		\end{eqnarray}
		
		\section{Diagonalization of corrected $M_N^q$}
		\label{app.Rs}
		The mass matrix $M_N^q$ with hierarchy corrections in Eq. (\ref{eq.MNdelta}) is diagonalized into leading order eigenvalues $(h_{12}^qh_{23}^q, h_{23}^q,1-h_{23}^q)$ by transforamtion $R_D^q(\theta)$
		\begin{eqnarray}
			R_D^qM_N^q(R_D^q)^T=\textrm{diag}(h_{12}^qh_{23}^q, h_{23}^q,1-h_{23}^q).
		\end{eqnarray} 
		Generally, $R_D^q(\theta)$ can be expressed by eigenvector $v_i$
		\begin{eqnarray}
			R_D^q&=&\Array{c}{v_1,v_2,-v_3}
		\end{eqnarray} 
		Here, the column vector $v_i$ is
		\begin{eqnarray}
			v_i=N_i\Array{c}
			{\delta_{12} \delta_{23} + \delta_{12} l_2 + \delta_{23} l_1 l_2 - \delta_{13} l_2^2 + \delta_{13} E_i + l_1 E_i\\
				\delta_{12} \delta_{13} + \delta_{12} l_1 + \delta_{13} l_1 l_2  - \delta_{23} l_1^2 + \delta_{23} E_i + l_2 E_i\\
				-(\delta_{12}^2 + 2 \delta_{12} l_1 l_2 + l_1^2 E_i + l_2^2 E_i - E_i^2)
			}
		\end{eqnarray} 
		with normalized coefficient $N_i$ and $i-$th eigenvalues $E_i$ of $(h_{12}^qh_{23}^q, h_{23}^q,1-h_{23}^q)$.

		\section{Fit Data}
		\label{app.fitdata}
		\subsection{fit the CKM mixing}
		\begin{itemize}
			\item Pattern: $l_1=1,l_2=1$.  
			\\
			A fit point: $$\theta^u=2.610,~\theta^d=2.367,~\lambda_1=0.05426,~\lambda_2=0.1026.$$
			Fit results: $$s_{12}=0.2254,~s_{23}=0.04173,~s_{13}=0.003715,~\delta_{CP}=1.131.$$
			Reconstructed mass matrices:
			\begin{eqnarray*}
				M_N^u&=&\frac{1}{3}\Array{ccc}{1 & 1 & 1 \\ 1 & 1 & 1 \\ 1 & 1 & 1}
				+\Array{ccc}{-0.00001503 & -0.001848 & -0.001867 \\
					-0.001848 & -0.00002050 & -0.007356 \\
					-0.001867 & -0.007356 & 0.00004803},
				\\
				M_N^d&=&\frac{1}{3}\Array{ccc}{1 & 1 & 1 \\ 1 & 1 & 1 \\ 1 & 1 & 1}
				+\Array{ccc}{-0.0009518 & -0.003920 & -0.01024 \\
					-0.003920 & -0.00002050 & -0.02054\\
					-0.01024 & -0.02054 & 0.003082 }.
			\end{eqnarray*}
			\item Pattern: $l_1=\frac{1}{2},l_2=\frac{1}{2}$.
			\\
			A fit point: $$\theta^u=3.0904,~\theta^d=0.06419,~\lambda_1=-0.002627,~\lambda_2=0.1046.$$
			Fit results: $$s_{12}=0.2246,~s_{23}=0.04192,~s_{13}=0.003702,~\delta_{CP}=1.106.$$
			Reconstructed mass matrices:
			\begin{eqnarray}
				M_N^u&=&\frac{2}{3}\Array{ccc}{\frac{1}{4} & \frac{1}{4} & \frac{1}{2} \\
					\frac{1}{4} & \frac{1}{4} & \frac{1}{2} \\
					\frac{1}{2} & \frac{1}{2} & 1 }
				+\Array{ccc}{0.00000354 & -0.004176 & -0.00001205 \\
					-0.004176 & -0.00005625 & -0.009075 \\
					-0.00001205 & -0.009075 & 0.00006521},~~~
				\\
				M_N^d&=&\frac{2}{3}\Array{ccc}{\frac{1}{4} & \frac{1}{4} & \frac{1}{2} \\
					\frac{1}{4} & \frac{1}{4} & \frac{1}{2} \\
					\frac{1}{2} & \frac{1}{2} & 1 }
				+\Array{ccc}{0.007041 & -0.01884 & -0.0008883 \\
					-0.01884 & -0.002404 & -0.02707 \\
					-0.0008884 & -0.02707 & 0.002842}.
			\end{eqnarray}
			\item Pattern: $l_1=\frac{1}{2},l_2=\frac{1}{4}$.
			\\
			A fit point: $$\theta^u=3.059,~\theta^d=2.815,~\lambda_1=0.01368,~\lambda_2=0.1197.$$
			Fit results: $$s_{12}=0.2248,~s_{23}=0.04157,~s_{13}=0.003819,~\delta_{CP}=1.101.$$
			Reconstructed mass matrices:
			\begin{eqnarray}
				M_N^u&=&\frac{16}{21}\Array{ccc}{\frac{1}{4}& \frac{1}{8} & \frac{1}{2} \\
					\frac{1}{8} & \frac{1}{16} & \frac{1}{4} \\
					\frac{1}{2} & \frac{1}{4} & 1}
				+\Array{ccc}{-5.542\times10^{-6} & -0.007496 & -0.00006466 \\
					-0.007496 & -1.926\times 10^{-5} & -0.01624 \\
					-0.00006466 & -0.01624 & 3.731\times 10^{-5}},~~~
				\\
				M_N^d&=&\frac{16}{21}\Array{ccc}{\frac{1}{4}& \frac{1}{8} & \frac{1}{2} \\
					\frac{1}{8} & \frac{1}{16} & \frac{1}{4} \\
					\frac{1}{2} & \frac{1}{4} & 1}
				+\Array{ccc}{-0.0005135 & -0.01951 & -0.003987 \\
					-0.01951 & -8.471\times 10^{-5} & -0.05190 \\
					-0.003987 & -0.05190 & 0.001741}.
			\end{eqnarray}
			\item Pattern: $l_1=\frac{1}{3},l_2=\frac{1}{2}$.
			A fit point: $$\theta^u=3.062,~\theta^d=0.07245,~\lambda_1=0.001435,~\lambda_2=0.1053.$$
			Fit results: $$s_{12}=0.2251,~s_{23}=0.04296,~s_{13}=0.003650,~\delta_{CP}=1.136.$$
			Reconstructed mass matrices:
			\begin{eqnarray}
				M_N^u&=&\frac{36}{49}\Array{ccc}{\frac{1}{9} & \frac{1}{6} & \frac{1}{3} \\
					\frac{1}{6} & \frac{1}{4} & \frac{1}{2} \\
					\frac{1}{3} & \frac{1}{2} & 1}
				+\Array{ccc}{-1.428\times 10^{-6} & -0.002488 & -8.066\times 10^{-5}\\
					-0.002488 & 4.877\times 10^{-5} & -0.009249 \\
					-8.066\times10^{-5} & -0.009249 & 6.270\times10^{-5}},~~~
				\\
				M_N^d&=&\frac{36}{49}\Array{ccc}{\frac{1}{9} & \frac{1}{6} & \frac{1}{3} \\
					\frac{1}{6} & \frac{1}{4} & \frac{1}{2} \\
					\frac{1}{3} & \frac{1}{2} & 1}
				+\Array{ccc}{4.669\times10^{-4} & -0.01412 & -0.001541 \\
					-0.01412 & -0.002329 & -0.02835 \\
					-0.001541 & -0.02835 & 0.003005}.
			\end{eqnarray}
		\end{itemize}
		
		\subsection{fit the PMNS mixing}
		Neutrino hierarchies are set as $h_{23}^\nu=0.173,h_{12}^\nu=0.114$.
		\begin{itemize}
			\item Pattern: $l_1=1,l_2=1$.
			A fit point: $$\theta^e=1.783,~\theta^\nu=0.3792,~\lambda_1=1.856,~\lambda_2=1.627.$$
			Fit results: $$s^2_{12}=0.2937,~s^2_{23}=0.5770,~s^2_{13}=0.02149,~\delta_{CP}=1.351\pi.$$
			Reconstructed mass matrices:
			\begin{eqnarray}
				M_N^e&=&\frac{1}{3}\Array{ccc}{1 & 1 & 1 \\ 1 & 1 & 1 \\ 1 & 1 & 1}
				+\Array{ccc}{-1.739\times 10^{-4} & -0.006302 & -0.05681 \\
					-0.006302 & -1.531\times 10^{-4} & -0.02694 \\
					-0.05681 & -0.02694 & 6.146\times 10^{-4}},
				\\
				M_N^\nu&=&\frac{1}{3}\Array{ccc}{1 & 1 & 1 \\ 1 & 1 & 1 \\ 1 & 1 & 1}
				+\Array{ccc}{0.04149 & -0.1233 & -0.1037 \\
					-0.1233 & -0.1511 & -0.06626 \\
					-0.1037 & -0.06626 & 0.1294}.
			\end{eqnarray}
			\item Pattern: $l_1=1/2,l_2=1/2$.
			A fit point: $$\theta^e=2.072,~\theta^\nu=3.124,~\lambda_1=2.452,~\lambda_2=1.647.$$
			Fit results: $$s^2_{12}=0.3047,~s^2_{23}=0.5847,~s^2_{13}=0.02164,~\delta_{CP}=0.9587\pi.$$
			Reconstructed mass matrices:
			\begin{eqnarray}
				M_N^e&=&\frac{2}{3}\Array{ccc}{\frac{1}{4} & \frac{1}{4} & \frac{1}{2} \\
					\frac{1}{4} & \frac{1}{4} & \frac{1}{2} \\
					\frac{1}{2} & \frac{1}{2} & 1 }
				+\Array{ccc}{1.486\times 10^{-4} &  -0.002191 &  -0.05764 \\
					-0.002191 &  1.560\times 10^{-4} &  -0.03175 \\
					-0.05764 &  -0.03175 &  -1.697\times 10^{-5}},
				\\
				M_N^\nu&=&\frac{2}{3}\Array{ccc}{\frac{1}{4} & \frac{1}{4} & \frac{1}{2} \\
					\frac{1}{4} & \frac{1}{4} & \frac{1}{2} \\
					\frac{1}{2} & \frac{1}{2} & 1 }
				+\Array{ccc}{0.03573 & -0.07414 & -0.02683 \\
					-0.07414 & -0.02880 & -0.2941 \\
					-0.02683 & -0.2941 & 0.01279}.
			\end{eqnarray}	
		\end{itemize}
		Neutrino hierarchies are set as $h_{23}^\nu=0.01,h_{12}^\nu=0.01$.
		\begin{itemize}
			\item Pattern: $l_1=1,l_2=1$.
			A fit point: $$\theta^e=2.197,~\theta^\nu=2.687,~\lambda_1=1.948,~\lambda_2=1.727.$$
			Fit results: $$s^2_{12}=0.3122,~s^2_{23}=0.5465,~s^2_{13}=0.02308,~\delta_{CP}=1.279\pi.$$
			Reconstructed mass matrices:
			\begin{eqnarray}
				M_N^e&=&\frac{1}{3}\Array{ccc}{1 & 1 & 1 \\ 1 & 1 & 1 \\ 1 & 1 & 1}
				+\Array{ccc}{-2.624 \times 10^{-5} &  -9.809\times 10^{-4} & -0.03931 \\
					-9.809\times 10^{-4} &  -2.700\times 10^{-5} & -0.04981 \\
					-0.03931 &  -0.04981 &  3.408\times 10^{-4}},
				\\
				M_N^\nu&=&\frac{1}{3}\Array{ccc}{1 & 1 & 1 \\ 1 & 1 & 1 \\ 1 & 1 & 1}
				+\Array{ccc}{-7.372\times 10^{-5} & -0.003409 & -0.001811 \\
					-0.003409 & -1.133\times 10^{-4} & -0.009848 \\
					-0.001811 & -0.009848 & 2.870\times 10^{-4} }.
			\end{eqnarray}
			\item Pattern: $l_1=1/2,l_2=1/2$.
			A fit point: $$\theta^e=2.082,~\theta^\nu=1.557,~\lambda_1=1.743,~\lambda_2=1.833.$$
			Fit results: $$s^2_{12}=0.3233,~s^2_{23}=0.5187,~s^2_{13}=0.02197,~\delta_{CP}=1.100\pi.$$
			Reconstructed mass matrices:
			\begin{eqnarray}
				M_N^e&=&\frac{2}{3}\Array{ccc}{\frac{1}{4} & \frac{1}{4} & \frac{1}{2} \\
					\frac{1}{4} & \frac{1}{4} & \frac{1}{2} \\
					\frac{1}{2} & \frac{1}{2} & 1 }
				+\Array{ccc}{1.401\times 10^{-4}  & -0.001990 & -0.05701\\
					-0.001990 & 1.470\times 10^{-4} & -0.03244\\
					-0.05701 & -0.03244 & 5.158\times 10^{-7} },
				\\
				M_N^\nu&=&\frac{2}{3}\Array{ccc}{\frac{1}{4} & \frac{1}{4} & \frac{1}{2} \\
					\frac{1}{4} & \frac{1}{4} & \frac{1}{2} \\
					\frac{1}{2} & \frac{1}{2} & 1 }
				+\Array{ccc}{-3.683\times 10^{-5} & -0.004278 & -0.01259\\
					-0.004278 & -2.143\times 10^{-5} & -5.350\times 10^{-4} \\
					-0.01259 & -5.350\times 10^{-4}  & 1.583\times 10^{-4} }.
			\end{eqnarray}	
		\end{itemize}

	\end{appendix}

	
\end{document}